\documentclass[preprint2]{aastex}

\def\delalpha{{\Delta \alpha \over \alpha}}
\def\dellamcal{\Delta\lambda_{cal}}
\def\etal{et al.\ }

\def\ten#1{\times 10^{#1}}
\def\aten#1{10^{#1}}

\def\chidof{\chi^2/{\rm d.o.f}}
\def\vcor{v_{\rm cor}}

\def\dellamcal{\Delta \lambda_{\rm cal}}

\def\deltatin{$\Delta {\rm Tempin}$}

\begin{document}

\title{Wavelength Accuracy of the Keck HIRES Spectrograph and Measuring Changes in the Fine Structure Constant}

\author{      
      Kim Griest\altaffilmark{1},
      Jonathan B.~Whitmore\altaffilmark{1},
      Arthur M. Wolfe\altaffilmark{1,2},
      J.~Xavier Prochaska\altaffilmark{2,3,4},
      J.~Christopher Howk\altaffilmark{2,5},
      and Geoffrey W.~Marcy\altaffilmark{6}
      \\
      }

\altaffiltext{1}{Department of Physics, University of California,
    San Diego, CA 92093, USA.}
\email{\tt kgriest@ucsd.edu, jonathan.b.whitmore@gmail.com}
\altaffiltext{2}{Visiting Astronomer, W.M.~Keck Telescope.  The Keck Observatory
is a joint facility of the University of California and the California 
Institute of Technology.}
\altaffiltext{3}{University of California Observatories - Lick Observatory, 
University of California, Santa Cruz, CA 95064, USA.}
\altaffiltext{4}{ Department of Astronomy and Astrophysics, University of 
California, Santa Cruz, Santa Cruz, CA 95064, USA.}
\altaffiltext{5}{Department of Physics, University of Notre Dame, Notre Dame, 
IN 46616.}
\altaffiltext{6}{Department of Astronomy, University of California, Mail Code 3411, Berkeley, CA 94720, USA.}

\begin{abstract} 
We report on an attempt to accurately wavelength calibrate four nights of data
taken with the Keck HIRES spectrograph on QSO PHL957, for the purpose of
determining whether the fine structure constant was different in the past. 
Using new software and techniques, we measured the redshifts of 
various Ni II, Fe II, Si II, etc.  lines in a damped Ly$\alpha$
system at $z=2.309$.  Roughly half the data were taken through the Keck iodine
cell which contains thousands of well calibrated iodine lines.  Using
these iodine exposures to calibrate the normal Th-Ar Keck data pipeline output 
we found absolute wavelength offsets of 500 m/s to 1000 m/s 
with drifts of more than 500 m/s over a single night, and drifts of nearly
2000 m/s over several nights.  These offsets correspond to an absolute
redshift of uncertainty of about $\Delta z \approx \aten{-5}$
($\Delta \lambda \approx 0.02$\AA), 
with daily drifts of around $\Delta z \approx 5\ten{-6}$ ($\Delta \lambda 
\approx 0.01$\AA), and multiday drifts of nearly $\Delta z \approx 2\ten{-5}$ 
($\approx 0.04$\AA).
The causes of the wavelength offsets are not known, but
since claimed shifts in the fine structure constant would result in
velocity shifts of less than 100 m/s,
this level of systematic uncertainty may make it difficult to use 
Keck HIRES data
to constrain the change in the fine structure constant.  
Using our calibrated data, we applied both our own fitting software
and standard fitting software to measure $\delalpha$, but discovered that we
could obtain results ranging from significant detection of either sign,
to strong null limits, depending upon which sets of lines and which fitting
method was used.  We thus speculate that the discrepant results 
on $\delalpha$ reported in
the literature may be due to random fluctuations coming from under-estimated
systematic errors in wavelength calibration and fitting procedure.
\bigskip
\end{abstract}

\vskip .2truein
\bigskip

\section{Introduction: Varying Fine Structure Constant}

The fine structure constant today, $\alpha_0 = 1/137.03599911$, is usually
thought of as a fundamental, unchanging, constant of nature, but recently
both experimental and theoretical papers have challenged this assumption,
(e.g., see the review by Garcia-Berro, Isern, and Kubyshin (2007)).
Motivated especially by the possible experimental detection (see below)
of a change
in $\alpha$, we applied for and took four nights of Keck data, half
of it through the Keck iodine cell, on QSO PHL957.  
Our goal was to get an extremely 
well-calibrated, high signal/noise absorption
spectrum on a distant damped Ly$\alpha$ (DLA) system ($z=2.309$) in 
order to measure the 
value of the fine structure constant more than 10 billion years ago and 
compare it to the value today.  

The basic method to determine $\delalpha = (\alpha_z-\alpha_0)/\alpha_0$,
is to measure differences in redshifts between different atomic
transitions of elements in the same physical system, and
use the fact that for small changes in $\alpha$
the energy level of a given atomic transition can be approximated as
\begin{equation}
\omega_\alpha = \omega_0 + 2 q \delalpha,
\label{eqn:eqdelalpha}
\end{equation}
where $\omega_0=1/\lambda_0$ is the frequency of the transition on Earth, 
$\omega_\alpha$ is the frequency in the high redshift cloud,
and the $q$-values measure the dependence of $\omega_\alpha$ 
on $\alpha$ and have been calculated 
for many common transitions.
(see for example, Murphy et al. 2001a, Dzuba, et al. 2002, and
Porsev, et al. 2007 for 
more detailed discussion and Table~\ref{tab:lineinfo}
for the values of $q$ for 
various transitions.)
The values of $q$ depend upon the electron orbital configurations of
the initial and final quantum states, and therefore different
transitions have different values of $q$.
If all transitions had the same $q$, then all wavelength shifts would be the
same and one could absorb any change in $\alpha$ into the determination of
the redshift of the physical system.
However, since different transitions have different values of 
$q$, the relative transition wavelengths will 
differ from what they are in the lab if $\delalpha \neq 0$.

For example, 
Murphy, et al. (2001a; 2001b; 2003; 2004), used a many-multiplet 
method on Keck HIRES data of 143 absorption systems in 
the redshift range $0.2 < z < 3.7$ to find
a significant reduction of $\alpha$ in the past,
$\delalpha =  (-5.43 \pm 1.16) \ten{-6}$, while Chand, et al. (2004)
and Srianand et al (2004),
used the same method for a subset of transitions on VLT/UVES 
data on 23 absorbers to find
$\delalpha =  (-0.6 \pm 0.6) \ten{-6}$.  
This latter analysis was criticized
by Murphy, et al. (2008) who reanalyzed the Chand et al. data to get
$\delalpha = (-4.4 \pm 3.6) \ten{-6}$, consistent with their 
previous result.
In the meantime other groups returned results, for example, 
Levshakov, et al. (2006) used the VLT/UVES spectrograph
to study Fe II lines in one system at $z=1.15$
and found a null result, $\delalpha = (-.07 \pm 0.84) \ten{-6}$, and
in another system at $z=1.8$ to find 
$\delalpha = (5.4 \pm 2.5) \ten{-6}$ (Levshakov, et al. 2007).
Murphy, et al. (2006) have also criticized these results, claiming that
the data do not allow limits as strong as those reported.
Subsequently, the Levshakov, et al (2006) results were weakened to
$\delalpha = (-0.12 \pm 1.8) \ten{-6}$ for the $z=1.15$ system and 
$\delalpha = (5.7 \pm 2.7) \ten{-6}$ for the $z=1.8$ system (Molaro, et al.
2008), still a null result inconsistent with the detections.

Given the above inconsistent results, 
we were particularly interested in the Fe II $\lambda$1608 and 
Fe II $\lambda$1611
transitions since these have $q$ values that are both large and more
importantly of opposite sign.  Thus, if $\alpha$ was different in 
the past, the relative positions of these two lines should be 
significantly shifted from their laboratory values.  
For our DLA at $z=2.309$, a relative
shift between Fe II $\lambda$1608 and Fe II $\lambda$1611
of about $-136 \pm 21$ m/s is expected if the Murphy et al.
value $\delalpha = (-5.43 \pm 0.116) \ten{-6}$ is correct.  
(Fe II $\lambda$1608 shifts by $-54 \pm 12$ m/s, while Fe II $\lambda$1611 
shifts by $82 \pm 18$ m/s in the rest frame.)
Thus our goal was to centroid these lines to better than 50 m/s, so
as to determine $\delalpha$ in a single ion in
a single absorption system.  

Our method, which is close to that used by Levshakov, et al 2006,
contrasts with that of
Murphy et al. (2001a; 2001b; 2003; 2004) and Chand et al. (2004) where
the signal/noise was
not high enough to detect the $\delalpha$ signal in any single pair of lines;
they did a statistical averaging over many transitions in many absorption
systems, and thus might be subject to systematic errors in
selection, calibration, or averaging procedures.  
Since we expected to have superbly well-calibrated spectra, and these 
two Fe II lines appear in 
the same echelle order we hoped we could give convincing evidence for 
or against a change in the fine structure constant.

For our work, in addition to the Fe II lines, there are also several 
Ni II, Si II, Al II, and Al III lines that fall in 
the wavelength range covered by the iodine cell and that we can use.
In what follows, besides the Fe II $\lambda$1608/$\lambda$1611 pair, 
various other sets of lines are used.
Potentially we could fit all 16 lines that have calculated $q$ values
and that appear in our spectra:
Fe II $\lambda$1608/$\lambda$1611/$\lambda$2344, Ni II 
$\lambda$1709/$\lambda$1741/$\lambda$1751, Si II $\lambda$1526/$\lambda$1808,
Al III $\lambda$1854/$\lambda$1862, Al II $\lambda$1670, Cr II 
$\lambda$ 2062/$\lambda$2056/$\lambda$2066, and Zn II $\lambda$2026/$\lambda$2062.
If we restrict ourselves to lines that occur at wavelengths for which we have
iodine spectra we would use only the 9 lines:
Fe II $\lambda$1608/$\lambda$ 1611, NiII $\lambda$1709/$\lambda$1741/$\lambda$1751, 
Si II $\lambda$1808, Al III $\lambda$1854/$\lambda$1862, and Al II $\lambda$1670.  
If we worry that saturated lines (those with minimum flux less than 10\%) 
may not be accurately fit we can restrict 
ourselves to the 7 lines that meet the above criteria
and are unsaturated: FeII $\lambda$1611, NiII $\lambda$1709/$\lambda$1741/$\lambda$1751, 
SiII $\lambda$1808, and Al III $\lambda$1854/$\lambda$1862. 
Finally, we note that Al III has a systematically higher ionization potential 
than the other ions and is a sub-dominant ionization state of Al.  It thus
could exist in a physically different location.  Thus, we most reliably 
consider the five calibratable, unsaturated, singly ionized transitions: 
FeII $\lambda$1611, NiII $\lambda$1709/$\lambda$1741/$\lambda$1751, and SiII $\lambda$1808.

\section{Data and Extraction}
PHL957 is a bright (B=16.6) QSO at $z=2.7$, with a damped Ly alpha
system at $z=2.309$ (Beaver, et al 1972).
We obtained 5.5 hours of data on November 1, 2002, 
4 hours on Oct 3, 2004, 4 hours on Oct 4, 2004, and 5 hours on Oct 5, 2004.
Five of the 11 exposures taken in 2002 had the iodine cell in place,
while 6 of the 13 exposures taken in 2004 were taken through the iodine cell.
Table~\ref{tab:journal} shows how the iodine cell exposures were 
interspersed with the
non-iodine cell exposures. Table~\ref{tab:journal} also shows the times that the
relevant Th/Ar calibration arcs were taken, as well the temperatures inside
the HIRES enclosure, and whether or not the echelle gratings were 
moved between the
the QSO exposures and the relevant Th/Ar exposures.
The data from 2002 were acquired through the C1 decker 
(FWHM~$\approx 6 \, {\rm km \, s^{-1}}$),
with the kv380 blocking filter in place.  The data from 2004 were also 
observed through the C1 decker but with the kv418 blocking filter, and 
the CCD mosaic was binned by two in the spatial dimension.

The 2002 spectra were extracted and combined using MAKEE (Barlow 2002).
The 2004 spectra were extracted and 
combined using the XIDL\footnote{http://www.ucolick.org/$\sim$xavier/IDL}
package 
HIRedux\footnote{http://www.ucolick.org/$\sim$xavier/HIRedux/HIRES\_doc.html} (Bernstein, et al. in preparation).
The 2002 and 2004 data cannot be easily combined since a new 
mosaic of CCD's was installed
in the spectrograph in between these runs.  
We note that the 2002 data have a substantially worse signal/noise per pixel
(42 for the non-iodine cell exposures) compared to the combined 2004 
non-iodine cell data which have a S/N of about 70 per 
pixel.

\subsection{Iodine Cell}
The Keck iodine cell has been used extensively in searches for extra solar
planets using the Doppler technique (Butler, et al. 1996).
The cell is placed in the beam and superposes several narrow absorption
lines per Angstrom on the QSO spectrum between 4950\AA\ and 5900\AA.
As pointed out by Murphy, et al. (2001b), the different optical paths of the Th-Ar 
lamp and the QSO spectrum are a possible source of systematic error in the wavelength
calibration.  Using the iodine cell, 
wavelength calibration errors can be dramatically reduced since issues such
as atmospheric dispersion, guiding errors at the slit, and all changes to
the optics of the spectrograph are shared by both the iodine lines as well
as the QSO spectra.  In addition, the Th-Ar lamp spectra are done
at different times than the QSO spectra.

\begin{figure}
\epsscale{1.0}
\plotone{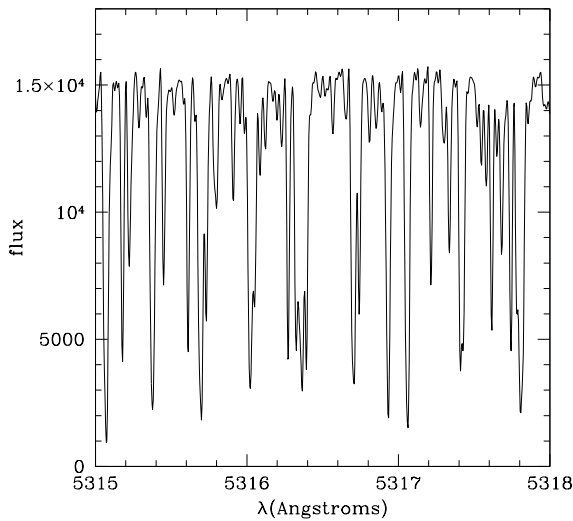}
\caption{Three Angstroms of the KPNO FTS spectrometer iodine cell spectrum.
\label{fig:iodine}
}
\end{figure}

\begin{figure}
\epsscale{1.0}
\plotone{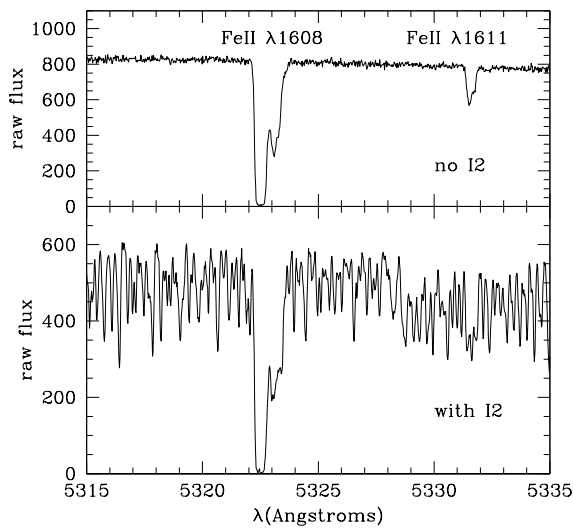}
\caption{
Combined 2004 absorption spectra showing a portion of order 67 containing the
Fe II $\lambda$1608 and FeII $\lambda$1611 lines. The top panel shows the spectrum taken without
the iodine cell in place, and the bottom panel shows the spectrum taken 
through the iodine cell.
\label{fig:fe67} 
}
\end{figure}

Figure~\ref{fig:iodine} shows a sample of the 
iodine cell spectra taken with the Fourier Transform Spectrometer (FTS)
at KPNO with a resolution of around 170,000 and a signal/noise of 
700 per pixel (Butler, et al., 1996; Marcy, 2008).  This high resolution and
S/N means that we do not expect the iodine spectrum to be a significant source
of wavelength calibration error.
Figure~\ref{fig:fe67} shows a portion of echelle order 67 taken
both with and without the iodine cell.
In the bottom panel of Figure~\ref{fig:fe67} one can see the iodine lines 
as well as the Fe II
$\lambda$1608 and Fe II $\lambda$1611 lines of the DLA towards PHL957.
The top panel of Figure~\ref{fig:fe67} does not contain the iodine lines. 

\subsection{Wavelength Calibration}

In order to do the wavelength calibration we started from the XIDL 
Th-Ar calibration.
\footnote{As a check we also extracted spectra using MAKEE, both with and
without the sky line wavelength calibration option.  In both cases iodine 
re-calibration of about the same magnitude as with XIDL was needed.}
XIDL takes Th-Ar wavelengths that have been transformed 
from vacuum values to air values so that lines can be identified.  
Using these identified lines a polynomial is fit over the full 2D spectrum.  
The wavelength scale found is then transformed back to vacuum values using 
the inverse Edlen formula.

After finding the Th-Ar wavelength scale we use our iodine measurements with
two similar but independent methods to recalibrate.  
In both methods we convolved the high signal/noise iodine spectrum
measured at KNPO (Butler, et al. 1996) with a Gaussian and then minimized 
the $\chi^2$ of the difference between this convolved spectrum and 
the PHL957 spectrum shifted by an amount $\dellamcal$.  

In the first method a 3-parameter fit was performed in each
wavelength bin of the PHL957 spectra, 
with bin sizes from 1 to 10 Angstroms being used.
The fit returned the wavelength shift, $\dellamcal$, the sigma of the
convolution Gaussian, as well as a multiplicative continuum offset 
and the formal errors of these quantities.
Using this method, regions of the PHL957 spectra that had strong lines 
were not included in the fit since
these could distort the results; a linear interpolation of bins on either 
side of the line region was used to find a correction at the
line center.

Using the second method, strong lines in the PHL957 spectra were fit and 
subtracted from the data before differencing with the convolved iodine
spectra.  This allowed direct determination of a wavelength correction
even at line centers (except for saturated regions which are removed as
above).
Also, in the second method rather than fitting for both $\dellamcal$ and 
the convolution sigma in each wavelength bin, 
a rolling  5 to 10 Angstrom bin iterative method was used
that found the wavelength correction for each data point and also the one best 
convolution sigma for the entire echelle order.
We also tried setting the sigma of our Gaussian convolution
kernal to the resolution expected
from the physical setup of the telescope, $R=50,000$.  
This sigma was consistent with but slightly smaller than the sigma we found 
by fitting, but 
the wavelength offsets found either way agreed to within errors.
In this method the errors in the wavelength calibration 
were found from the values of $\dellamcal$ that caused the
$\chi^2$ to increase by 1.

The methods of finding the continuum of the FTS iodine spectra
were also somewhat different for the two methods.
In the first method, we used the highest flux value in a variable 
size wavelength bin as the continuum value.  
In the second method, we averaged
the three largest flux values (and their corresponding wavelengths) in each
one Angstrom bin, and then set the continuum 
by fitting a spline through these averaged points.  

Determining the continua of the PHL957 spectra taken with the 
iodine cell is difficult because of the large number of unresolved
overlapping iodine lines.  The first method used 
a 4th order polynomial fit to the set flux values within 2-sigma of
the highest flux found over a certain wavelength range.  The second method
found the PHL957 continuum using a standard continuum fitting 
program, thus probably underestimating it.
Because of this underestimate, 
we repeated our analysis using several different possible continua and 
discovered that our final wavelength calibration results were robust for all 
plausible PHL957 continua.

A comparison between the results of the two methods showed agreement in
$\dellamcal$ within errors, and also good agreement in the reported errors
on $\dellamcal$.
We also tried a method that used a cross-correlation between the FTS and PHL957
spectra, which also worked, but did not seem as accurate as either
of the other two methods.  

We will use the results of our second method throughout the rest
of the paper.  Our final recalibration shift data is smoothed with 
a 5-Angstrom box filter to reduce fitting noise.

\begin{figure}
\epsscale{1.1}
\plotone{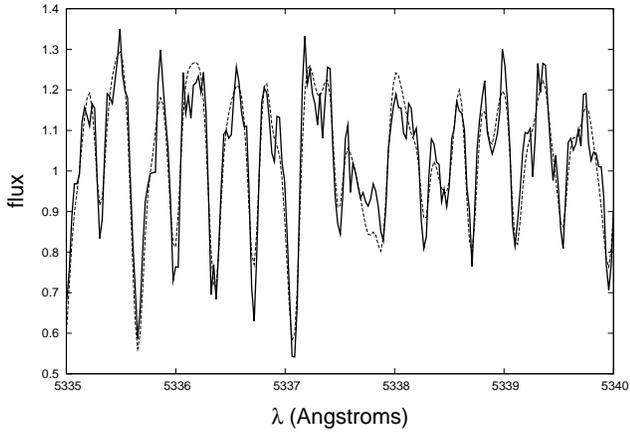}
\caption{
A small portion of echelle order 67
of PHL957 (thick line) compared with the convolved and shifted FTS 
iodine cell spectrum (thin dashed line).
\label{fig:convolve} 
}
\end{figure}

Figure~\ref{fig:convolve} compares a small portion of echelle order 67
of PHL957 with the convolved and shifted FTS iodine cell spectrum.

Note that the Th-Ar calibrated spectra used as input to our recalibration
pipeline should not include the standard Keck HIRES  heliocentric 
correction that 
accounts for the Doppler shift from the changing motion of the Earth 
around the Sun.  We recalibrate in the lab frame, and then make our
own, more accurate, solar system barycentric motion correction using
the code developed by Marcy and Butler (2008).  This barycentric
correction should be good to better than 1 m/s.

We found that the accuracy of the wavelength recalibration as measured by
our calculated error depended strongly on the size of the bin used for
the convolution.  Thus we report values from a 10\AA\ bin analysis
as a compromise between small uncertainty and constancy of the 
wavelength calibration over the width of a single bin.

\begin{figure}
\epsscale{1.1}
\plotone{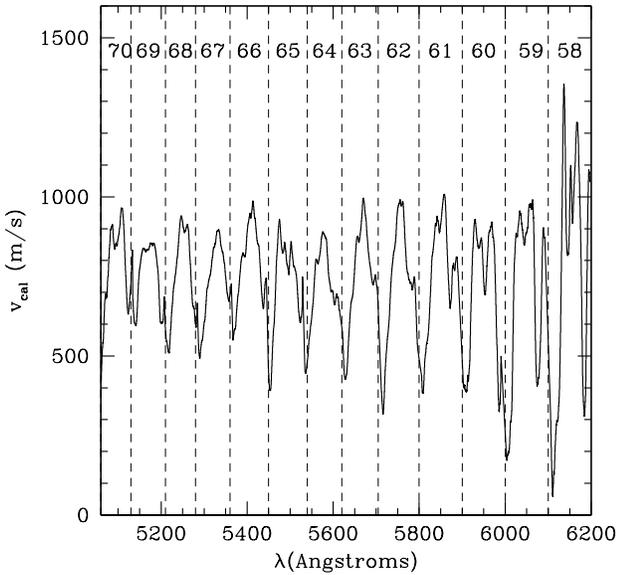}
\caption{
Wavelength correction over entire wavelength range for exposure 3-1 of
PHL957. The shift is between the standard Th-Ar wavelength 
calibration and the wavelength scale found by fitting to
the FTS iodine spectrum.  
The echelle order is marked.
\label{fig:vshiftall} 
}
\end{figure}

Figure~\ref{fig:vshiftall} shows an example (exposure 3-1) of the resultant
wavelength recalibation shift over the entire wavelength range for which we
have both QSO and FTS iodine spectra.  
This figure is a major result of this work.

\begin{figure}
\epsscale{1.1}
\plotone{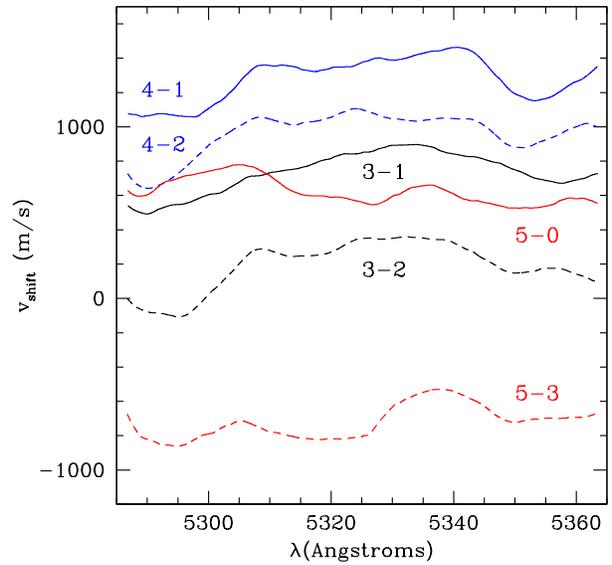}
\caption{
Time evolution of iodine cell wavelength recalibration shift 
for echelle order 67.  
Each line is labeled by the day-observation number, with the solid line
being the exposure taken earlier in the night.
\label{fig:vshift67}
}
\end{figure}
We see several interesting effects.  First the size of the shift is 
significant: typically from 500 m/s to 1000 m/s corresponding to 
a substantial fraction of a CCD pixel (roughly 1300 m/s/pixel or 
0.023\AA/pixel at $\lambda \approx 5300$\AA.)
Contrast these systematic velocity errors with an expected shift of 
136 m/s between Fe II $\lambda$1608 and
Fe II $\lambda$1611 due to the $\delalpha$ claim described in the introduction.
We note previous workers have found similar Keck HIRES shifts of 
1000m/s or larger 
(e.g. page 734 of the night airglow line paper by
Osterbrock et al. 2000), and Figure 4 of Suzuki et al. 2003)
but this does not seem to be widely appreciated.

Second, there is a clear pattern seen in each echelle order, with the 
shift increasing from the edge of each order and reaching a maximum
near the middle. 
For the purpose of measuring $\delalpha$ it is not the overall shift discussed
above that is important, but the relative shift between the transitions being
compared.  Depending on the echelle order Figure~\ref{fig:vshiftall} shows 
relative shifts of 300 m/s to 800 m/s within the same order.  These shifts
could be dangerous since, depending upon the lines being being compared,
they could result in a systematic relative velocity shift between lines,
thereby mocking a changing $\alpha$.

Next, we are interested in how this wavelength recalibration shift
varies with time. 
Thus, Figure~\ref{fig:vshift67} shows a more
detailed look at echelle order 67 which contains the
Fe II $\lambda$1608/$\lambda$1611 lines.  
In the figure,
different lines are the recalibration shifts for each of the five exposures
that used the iodine cell and are labeled by their ID's, 
e.g. 3-0 is the first exposure of PHL957 taken on Oct-3-2004.
See Table~\ref{tab:journal} for more details.

We note that each exposure has a similar, but not identical shape as a 
function of wavelength, and that there is a different wavelength
offset for each exposure.  While the variation over this order for
each exposure is typically 500 m/s or less, the shift between different
exposures can be as much as 2000 m/s.
This means that the Th-Ar wavelength calibration that gives us the
wavelength solution as a function of pixel, is not stable and drifts with time.

It is of interest to note that recalibration offsets between nights
are in general larger than the drifts during each night.  
One sees that on Oct 3 and 4, when the iodine cell exposures where
taken one after the other, the shift during the night was substantially
smaller than the inter-day shifts, while on Oct 5, when there was 
three hours between iodine cell exposures, there was a larger shift.  

Figure~\ref{fig:vshift67} is probably more important than
the previous figure, since it shows that the Th-Ar calibration is
not stable over time.  If the wavelength shifts shown in 
Figure~\ref{fig:vshiftall} were stable in time,
they could be removed and would have little effect on the measurement of 
$\delalpha$.  But large systematic wavelength shifts during the night mean that
measurements at the desired level of precision may be difficult
with Th-Ar calibration alone.  

It is important to note that even though
the calibration errors reported here are much larger than the 
final velocity precision needed to determine $\delalpha$, it is possible that 
these
calibration errors average out and do not ruin the final $\delalpha$
determination.  In the many multiplet method, many different lines are
compared across many different absorbers at many different redshifts.  
If the signs of the calibration errors
are random, the calibration errors may average away.  A complete discussion
of this possibility is beyond the scope of this work, but will be presented
elsewhere.

\subsection{Understanding the Calibration Shifts}
We made some preliminary attempts at understanding the causes 
of these unexpected
wavelength calibration shifts.  While complete understanding of the causes
and exploration of methods to correct for the shifts are beyond the scope
of this paper, we report some ideas and preliminary results here.  
We hope to use additional data and analysis to find a more complete 
understanding, which will then be published elsewhere.

First, it is interesting that during all three nights the shifts became
more negative at about 500 m/s per hour.  Thus we plotted the shifts vs time,
and also vs. the various temperatures, etc. that HIRES reports.
With only six iodine exposures it is difficult to discern a pattern
and impossible to that prove a pattern exists, and in fact we did not see any
fully convincing trends.

\begin{figure}
\epsscale{1.0}
\plotone{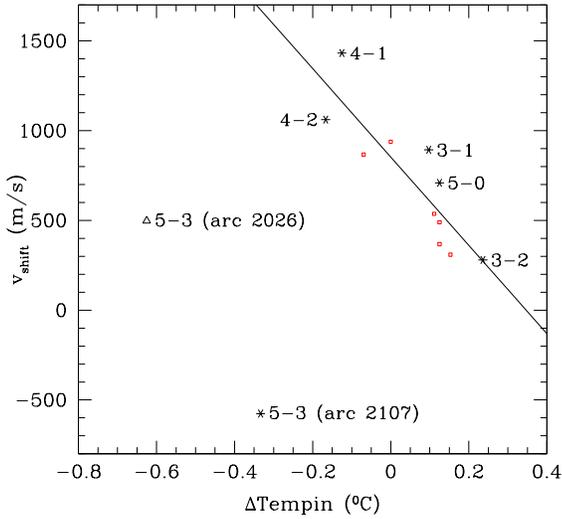}
\caption{Iodine recalibration shift $v_{\rm shift}$ vs \deltatin\ for the 
PHL957 iodine exposures (labeled asterisks), and 
a star HD209837 (small squares).
The shift is the single best wavelength recalibration shift for echelle
order 66 found by our iodine line fitting program.  The quantity \deltatin\
is the temperature inside the HIRES enclosure measured during the Th/Ar
arc calibration exposure minus the same temperature measured during the data
exposure.  The solid trend line, $v_{\rm shift} = -2459$ \deltatin $ + 853$.
is a fit to the points excluding QSO exposure 5-3.  Exposure 5-3 appears twice
labeled by which Th/Ar calibration arc was used. 
\label{fig:deltatin}
}
\end{figure}

The best partial trend is a decrease in calibration shift with \deltatin, where
\deltatin\ is the difference in HIRES enclosure temperature between when 
the Th/Ar calibration exposure 
was taken and when the QSO exposure was taken.  We plot this trend in
Figure~\ref{fig:deltatin}, where the 6 QSO exposures are shown as asterisks
and individually labeled.  The data is given in Table~\ref{tab:journal}.
We see that the trend is badly broken by the 4th exposure 
on Oct 5 (5-3 or ID 2097), which originally led us to discount temperature
as the main culprit.  However, we have some exposures of a star (HD209833) 
taken the same nights through the iodine cell, and we plot the calibration 
shifts vs \deltatin\ for them as small boxes.  
The complete analysis of these star exposures is beyond the
scope of this paper, but we see that they fall on the same trend line, 
showing that exposure 5-3 seems to be an exception.

Next, we check whether or not the echelle gratings were moved between the time
of the Th/Ar arc exposure and the QSO exposure.  This is shown in 
Table~\ref{tab:journal}.  We see that only on Oct 3 (exposures 3-1 and 3-2) were
the gratings not moved between Th/Ar calibration and the QSO exposures.  In 
Figure~\ref{fig:deltatin} we see that the dispersion of exposures 3-1 
and 3-2 from the trend line is smaller than for exposures 4-1 and 4-2.
Exposure 5-0 is near the trend line, but of course 
5-3 is way off, probably for some other reason.
Thus we see a weak hint that moving the gratings between Th/Ar calibration
and science exposure can cause a calibration error.

To understand the problem with exposure 5-3 we note that a different
Th/Ar arc was used for 5-3 than for the other iodine exposure (5-0) 
taken that night.  
To check whether something went wrong with the arc exposure (ID=2107) used to 
calibrate 5-3, we redid the iodine recalibration using arc exposure 2026,
taken much earlier that same night.
This point is also plotted in Figure~\ref{fig:deltatin} as a labeled triangle.
The point shifts substantially, but keeps the same
(large) distance from the trend line. 

Thus we see that while temperature surely plays an important role and 
a correction 
perhaps can made for this, there are other parameters that seem 
to be important, not all
of which are understood at this point.  Moving the gratings between Th/Ar
and science exposures may increase the calibration error. 
In what follows we will not attempt to
use the temperature trend to make any correction, though in future work
this might be possible.  

We note that other contributions to the calibration shifts are possible.
N.~Suzuki (private communication) suggested that the cause could be
that the HIRES spectrograph is mounted at a small angle with respect
to the optical axis, so that the light path rotates as the telescope moves, 
resulting
in variable vignetting in comparison to light from the Th-Ar lamp which
is fixed to the spectrograph.  More generally, the optical path of the
Th/Ar calibration light differs from the optical path of the QSO light
resulting in different wavelength to pixel mapping.  
P.~Molaro (private communication) suggested that the cause could be 
the changing position of the QSO image centroid in the slit.   We hope to
check this with additional exposures at a future date.
M.~Murphy (private communication) suggested that the cause could
be temperature and pressure differences between the Th/Ar calibration exposures
and the data exposures, as well as the resetting of the echelle gratings.
The analysis presented above was, in large part, motivated by the comments of 
M.~Murphy.

It would be useful to study and understand these wavelength calibration shifts,
perhaps by analyzing other data, since it might then be possible to model 
and remove them.  This will be pursued in more detail elsewhere.


Overall, the wavelength calibration errors reported here
may seem to be a remarkable result, but
we note that Osterbrock et al, (2000) recalibrated the Keck HIRES Th-Ar tube 
wavelengths using several night sky airglow lines and found similar
($\sim 0.05$\AA) calibration errors.   In addition, both the
magnitude and time variation of these shifts were detected
using Ly alpha forest lines by Suzuki, et al. (2003).

It is interesting to ask whether sky lines themselves could be used to
calibrate the QSO spectra without use of the iodine cell.  In fact, the current
version of MAKEE data reduction pipeline includes an option to calibrate using
sky lines.  
We used our iodine method to check on the wavelength accuracy of the MAKEE
sky line calibration, but still found substantial wavelength shifts both within
and across Echelle orders, and as a function of time.



\subsection{Use of the Calibration}
Our original hope was that if there would be a time-stable wavelength
calibration found from the iodine cell exposures then that then could be applied
to all the exposures.
When the iodine cell is in place we get well-calibrated wavelengths, but
substantially less S/N. 
This is because iodine lines cover basically the entire spectrum, decreasing 
the number of photons, adding spurious components to line fitting programs,
and also making the continuum difficult to determine.  So, it is the exposures 
without the iodine cell that carry the highest signal/noise and that we wish to
use to test for a changing alpha.
Since again, as Figure~\ref{fig:vshift67} shows, 
the wavelength calibrations vary substantially
over the course of a night, it is not clear how to use the calibration
in our fits.
We tried several methods before settling on the following:
1)  for exposures taken with the iodine cell we add the wavelength
recalibration correction to each wavelength output by the standard HIRES
pipeline.  2) For exposures taken without the iodine cell, we interpolate/
extrapolate the wavelength correction using the two iodine exposures nearest
in time, and assuming that the correction changes linearly with time.
3) We then add the barycentric correction to every data point.

At this point we have wavelength corrected spectra that can be fit.
We can either add the spectra for each order together (rebinning since
the spectra are no longer on a common wavelength scale) or just combine and
sort the data files for each order together giving more flux measurements
to be fit.

\begin{figure}
\includegraphics[angle=-90,scale=.3]{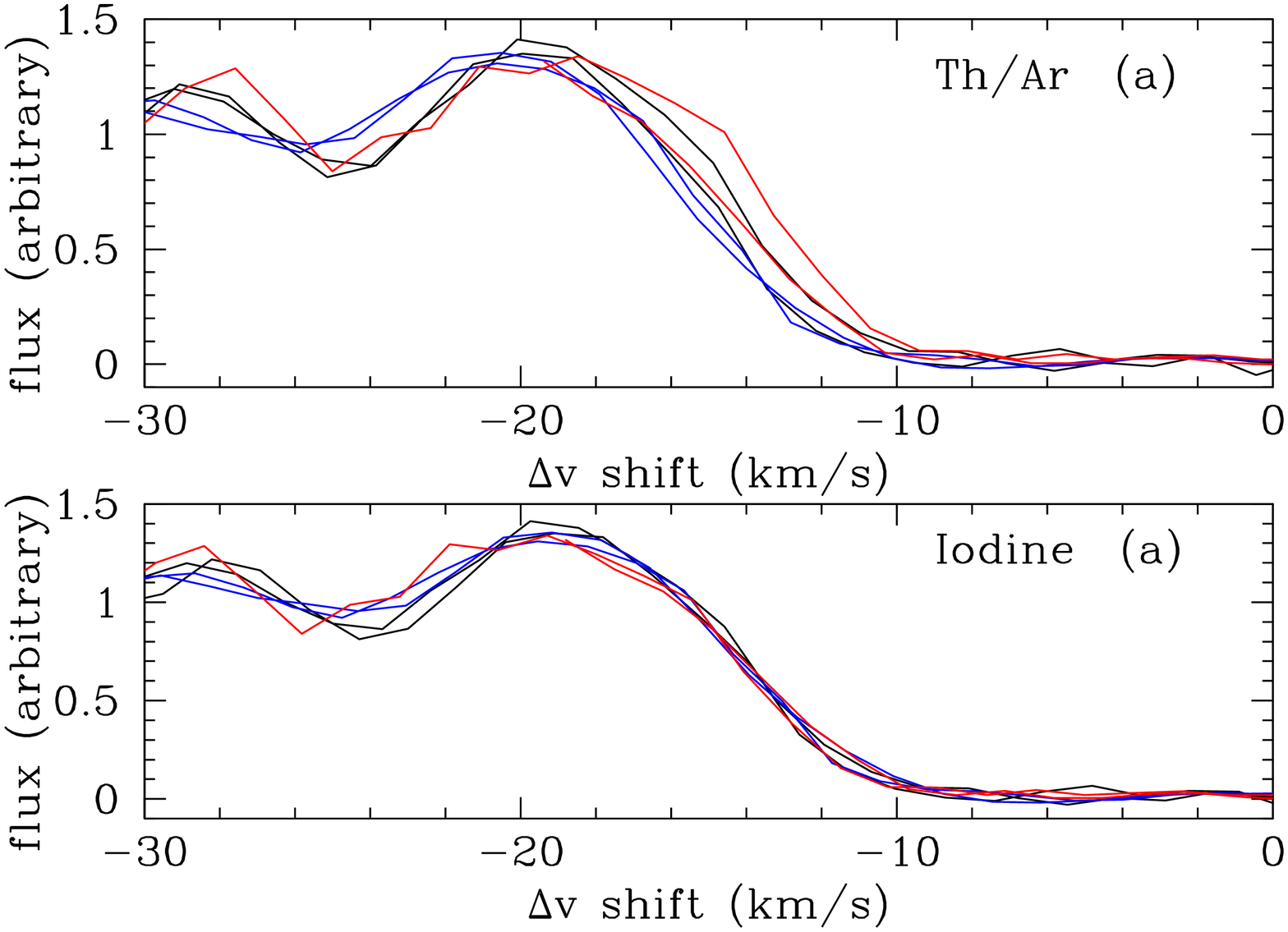}
\includegraphics[angle=-90,scale=.3]{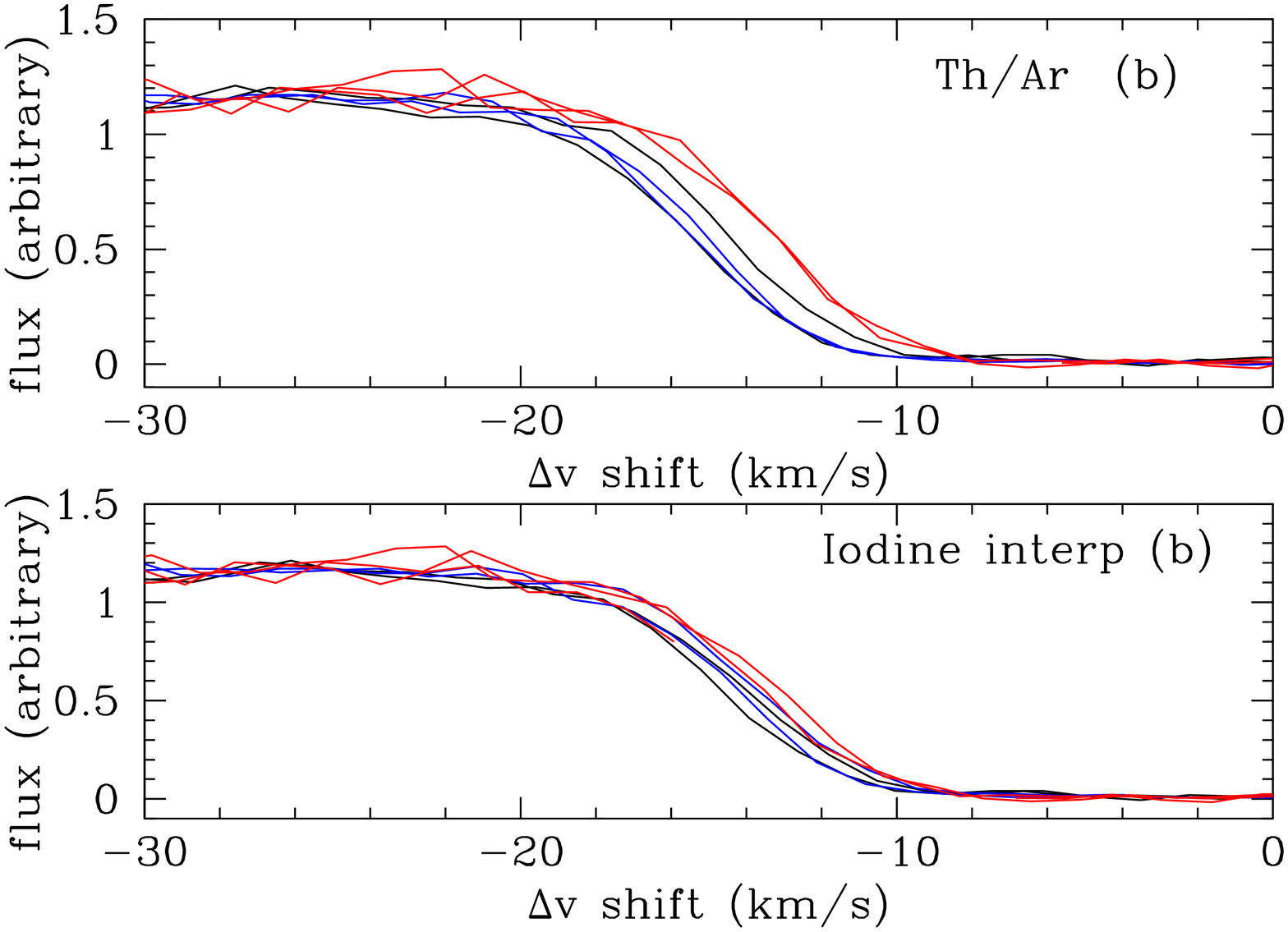}
\caption{
Close-ups of the left edge of the Fe II $\lambda$1608 line for observations
taken on October 3 (black), October 4 (blue), and October 5 (red).  
The panels labeled `Th/Ar' show results from the
standard Th/Ar wavelength calibration, 
while the panels labeled `Iodine', or `Iodine interp' show
results after recalibration with the iodine cell.
The two panels labeled (a) are for the 6 exposures with the iodine cell
in place,
while the two panels labeled (b) are for the 7 exposures taken without
the iodine cell, but recalibrated either using the iodine lines (upper panel)
or using interpolation as described in the text (lower panel).  The extra
bumps in the (a) panels are iodine lines, and the first part of the spectrum in
exposure 5-0 is missing due to a cosmic ray.  The iodine lines do not line up because
of the barycentric correction needed for PHL957.
\label{fig:checkvshift}
}
\end{figure}

We can check the effect of our recalibration by seeing how well
sharp features in the spectra line up.  For example,
Figure~\ref{fig:checkvshift} shows a close-up of the left edge of the
saturated Fe II $\lambda$1608 line before and after iodine recalibration.
The panel labeled `Th/Ar' (a) shows an overlay of the six 
iodine exposures calibrated with the standard HIRES pipeline, 
while the panel labeled `Iodine' (a) 
show the same after our recalibration.  The features line up
significantly better after recalibration;  spreads of more than 2000
m/s become significantly less than 1000 m/s.

The other two panels show overlays of the seven exposures taken without the
iodine cell.  The panel labeled `Th/Ar' (b) is again the result
from the standard HIRES calibration, while the panel labeled 
`Iodine interp' (b) shows the same seven exposures, but recalibrated using the 
interpolation scheme described above.
For the non-iodine exposures, 
there is some improvement in alignment of the line edge, but not nearly
as much as for the iodine exposures.  Here spreads of 1000 m/s
or more seem to remain even after recalibration.

Since for signal/noise and fitting reasons, we must use mainly
the non-iodine exposures, it is disappointing that our efforts to recalibrate
may not pay off.  Our inability to model the calibration shifts between
iodine exposures seems to be the main culprit.
~

%

\section{Fitting the Data}

In what follows, we mostly just combine the recalibrated, continuum subtracted
spectra for all non-iodine exposures into one large file, and then 
sort by wavelength and fit.   In some cases we rebin
the data combining several data points and adding the errors in quadrature,
and in some cases we just use the co-added spectra from the standard HIRES
pipeline.
We treat the data from 2002 separately since it was taken on a different
CCD chip.  We also treat the data with and without the iodine cell 
separately since the iodine lines add significant effective noise to 
the PHL957 spectra.


\subsection{Voigt profile fitting}
\begin{figure}
\includegraphics[angle=-90,scale=.3]{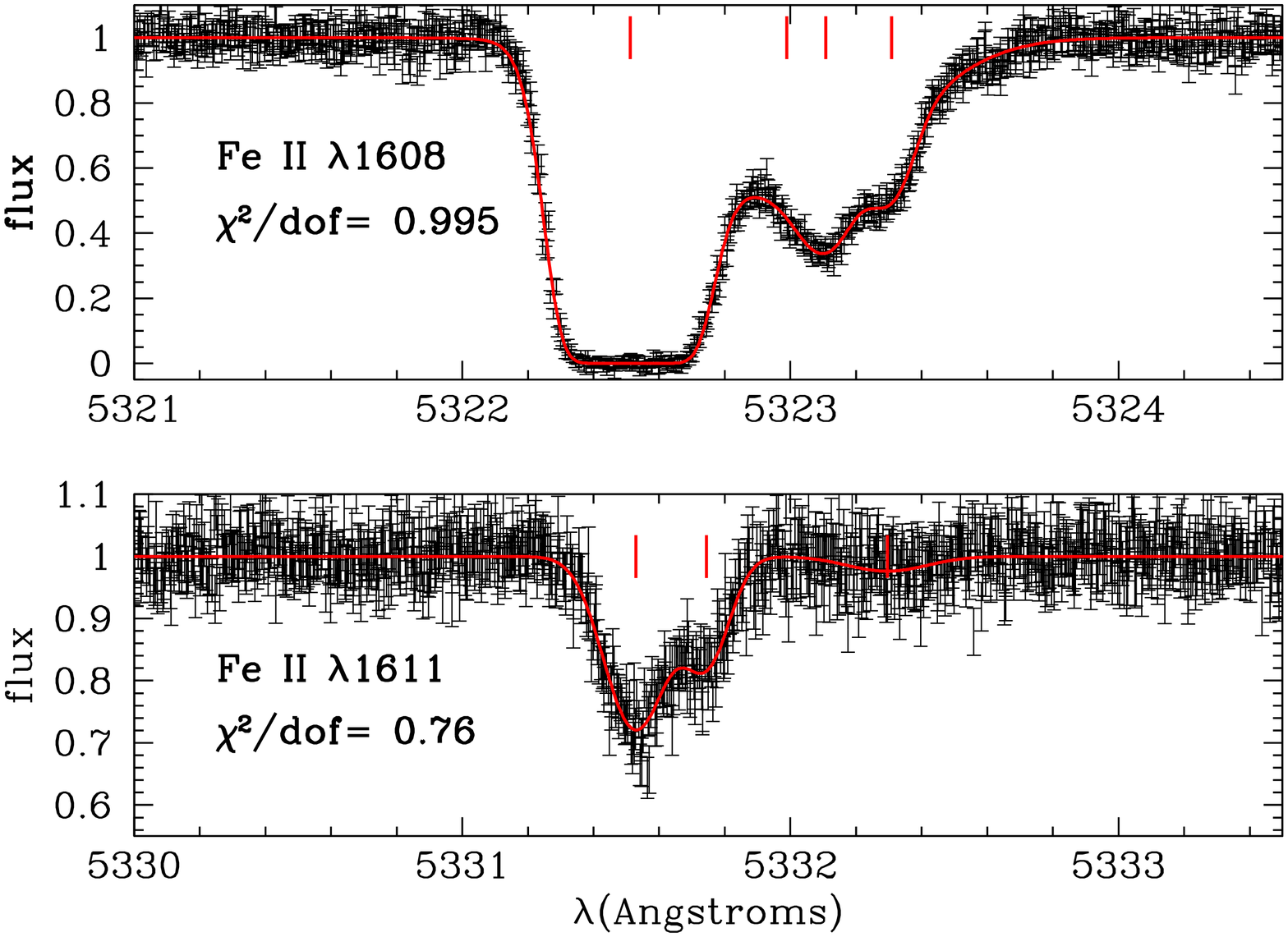}
\includegraphics[angle=-90,scale=.3]{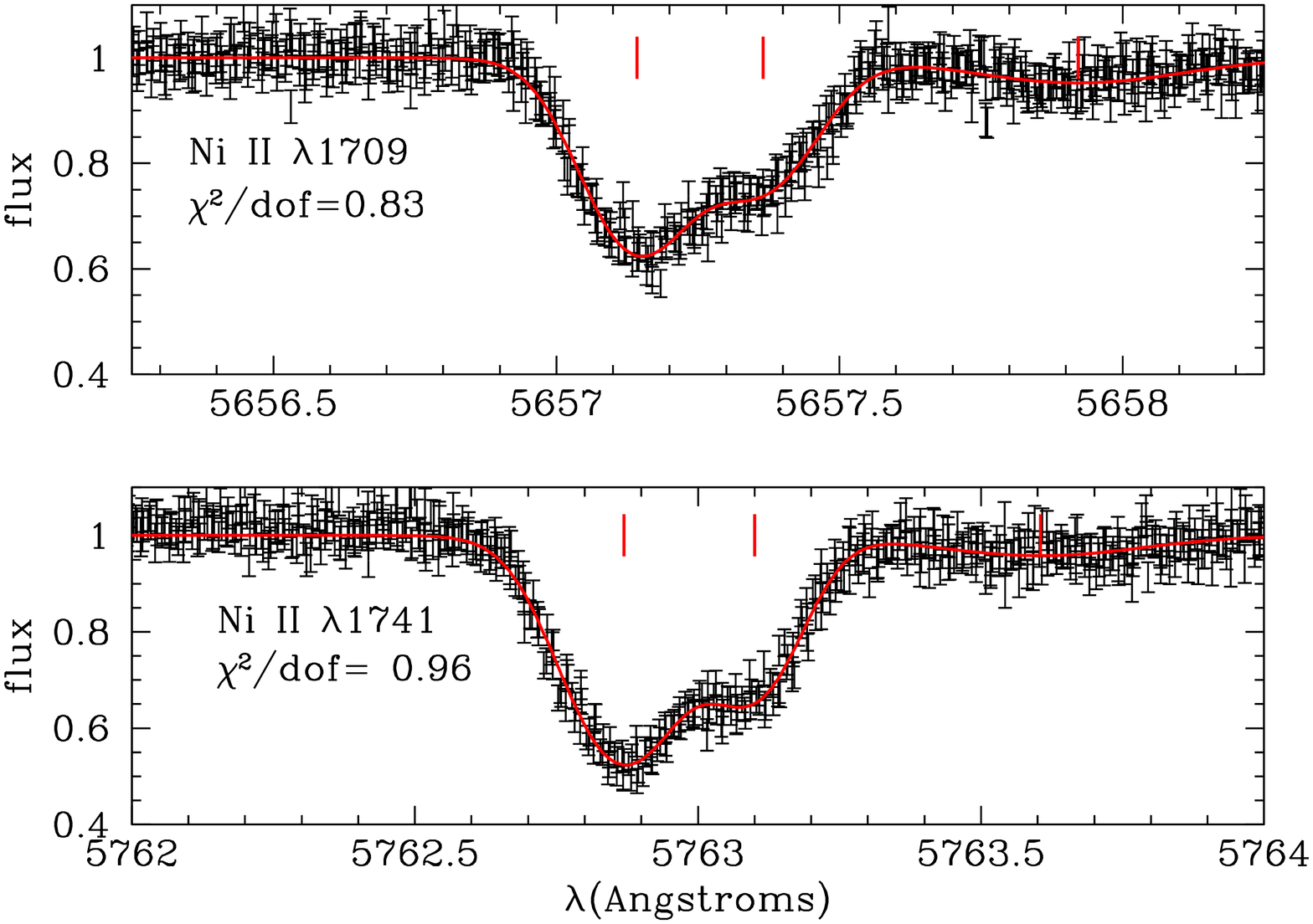}
\includegraphics[angle=-90,scale=.3]{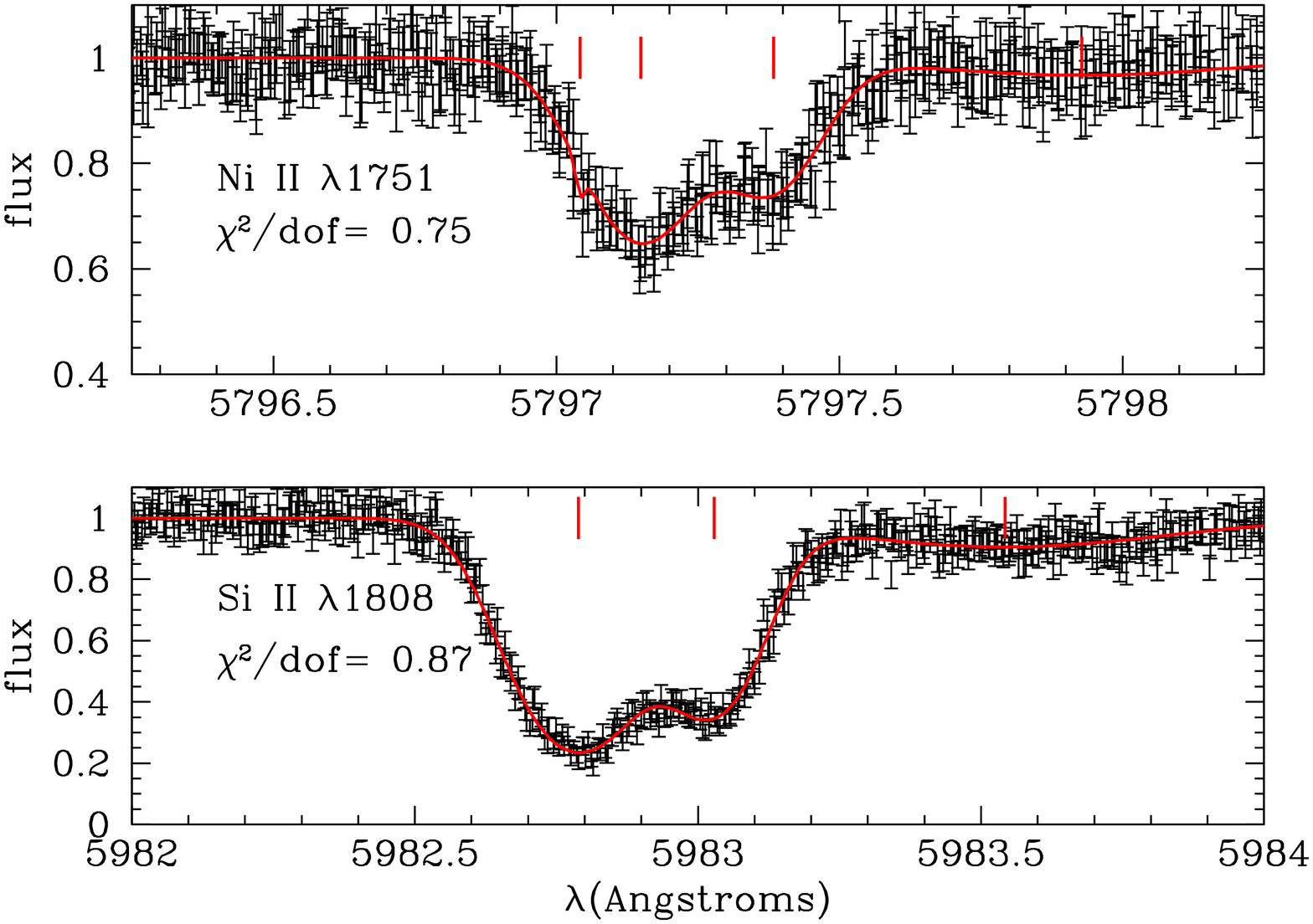}

\caption{
Voigt profile fit results for several PHL957 lines using our fitting code. 
The velocity components are marked.  These are unconstrained fits for each
line so the velocity components of various transitions are not forced to agree 
with each other in these plots.
\label{fig:linefita}
}
\end{figure}

\begin{figure}
\includegraphics[angle=-90,scale=.3]{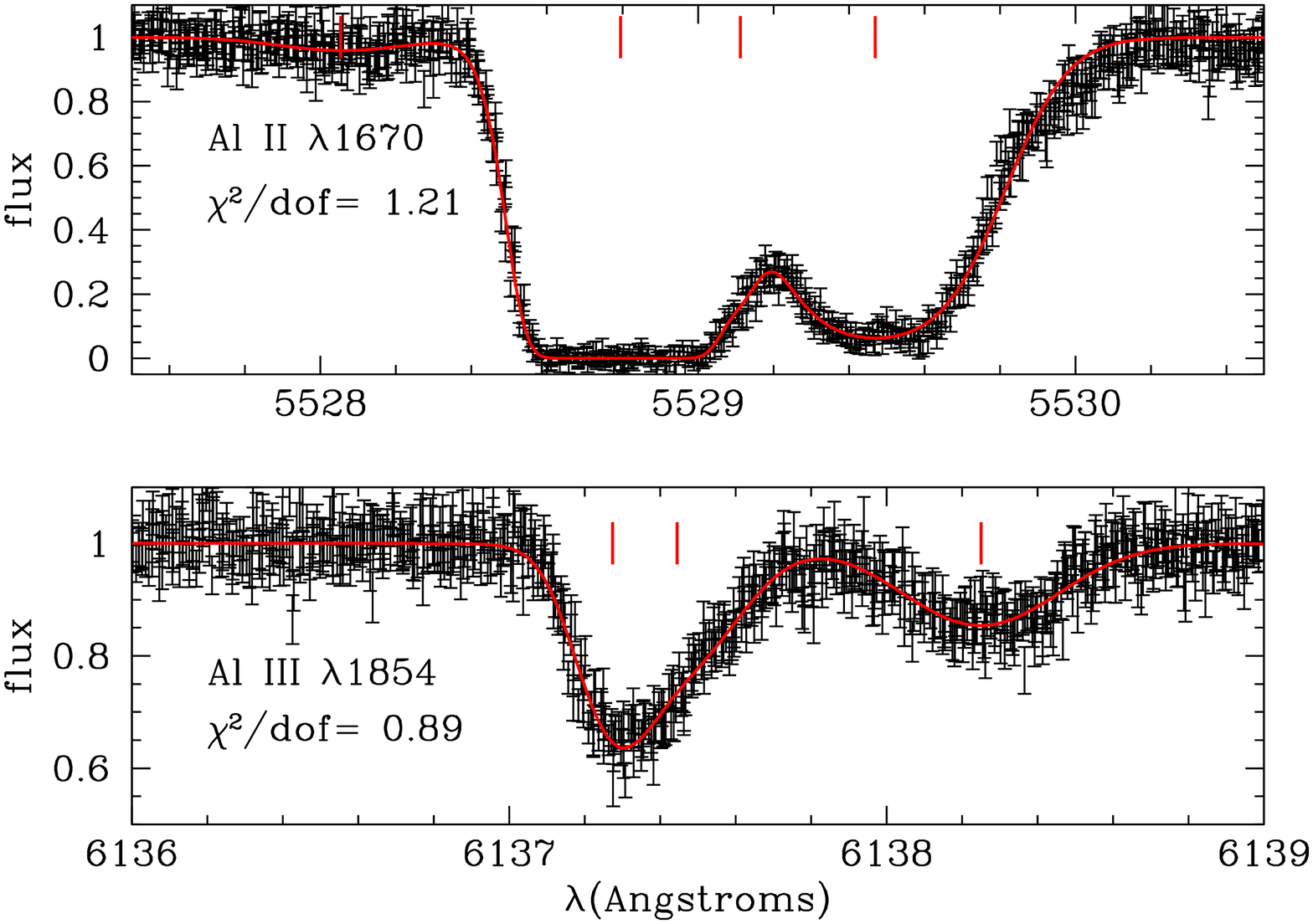}
\includegraphics[angle=-90,scale=.3]{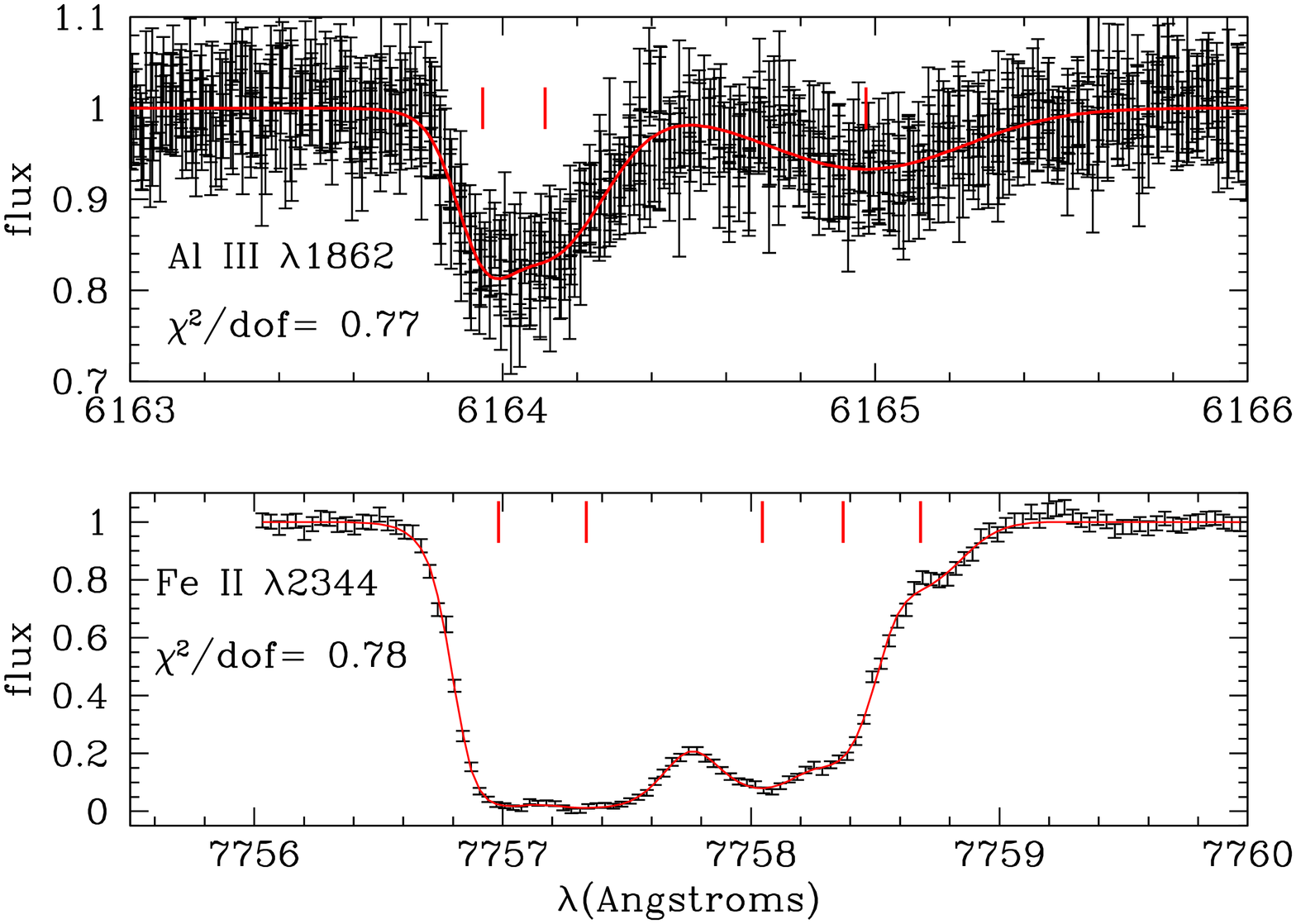}

\caption{
Voigt profile fit results for several PHL957 lines using our fitting code. 
The velocity components are marked.
\label{fig:linefitb}
}
\end{figure}

We fit the spectra using a code we developed based upon the CERN library MINUIT
minimization program and the humdev Voigt profile calculation 
routine (Wells, 1999).
We compared our results with those of 
VPFIT (Carswell et al. 2008) and DUDE (Kirkman et al. 2003) and found 
agreement for individual lines.
Using our own code allows us easily to do joint fits with
the redshifts of transitions varying independently and also allows us 
to add additional parameters as needed.
Figures~\ref{fig:linefita} and \ref{fig:linefitb}
show fit results for several transitions for the combined 2004 data.
The fit parameters are the redshift, the line width (b-value), and the column
density.  
We note that several of the lines are saturated, making wavelength centroiding
more difficult and less accurate.  

\subsection{Limit on precision of $\delalpha$}

The S/N of each 2004 non-iodine spectra is approximately 25 per pixel, giving
a total S/N for the 7 non-iodine exposures of about 70 per pixel.  
The individual iodine 
exposures have a S/N of around 19 per pixel, for a co-added total of about 
47 per pixel, though the iodine lines cause the effective S/N to
be lower than this. 
The co-added 2002 data have S/N approximately 42 per pixel.
Our total signal/noise is thus quite high for a high redshift object,
and we want to first estimate the minimum possible error on $\delalpha$
that could be obtained with these spectra.  

To do this, we use a Fisher matrix type method suggested by
Murphy, Webb, and Flambaum (2006), and Bouchy,  Pepe,  and Queloz (2001).
For continuum normalized flux spectrum $F(\lambda_i)$ with 1-sigma error
array $\sigma_F(\lambda_i)$, the minimum possible uncertainty in velocity
contributed by pixel $i$ is:
\begin{equation}
{\sigma_v(\lambda_i)\over c} = {\sigma_f(\lambda_i) \over \lambda_i 
(\partial F(\lambda_i)/\partial \lambda_i)}.
\label{eqn:eqsigmavi}
\end{equation}
Thus more precise velocity measurements come from pixels with large flux 
gradients and small errors.  The minimal possible uncertainty in the velocity
of a portion of spectrum is thus:
\begin{equation}
\sigma_v = \left[\Sigma_i [1/\sigma_v(\lambda_i)]^2\right]^{-1/2},
\label{eqn:eqsigmav}
\end{equation}
where the sum is over pixels.
Finally, the minimum uncertainty in $\delalpha$ can be found by performing
a least-squares fit to a version of Equation~\ref{eqn:eqdelalpha}:
\begin{equation}
v_j  = v_0 +\left(\delalpha\right) x_j,\qquad\qquad  x_j  = -2 c  q_j \lambda_{0j},
\label{eqn:eqvj}
\end{equation}
where $j$ numbers the lines that are being compared,
$v_0$ is a constant offset (degenerate with the system redshift),
and the minimum error in $\delalpha$ is just the fit uncertainty
in the slope of this linear equation.

Murphy, Webb, and Flambaum (2006) performed this analysis for their data 
and for the data of
Chand, et al. (2004), and Levshakov, et al. (2006), finding that 
while their own errors were (barely) within the allowable minimum, the
reported uncertainties of the others were smaller than the minimum possible.
The corrected version of the Levshakov results (Molaro, et al. 2008) 
does seem to be in agreement with the minimum error limit.

We would like to perform such an estimate, but first note that previous
estimates of minimum errors did not include uncertainties in the $q$ values.
Table~\ref{tab:lineinfo} shows
that these uncertainties can be significant, and inclusion of these uncertainties
will increase both the Fisher matrix minimum errors and also the error on
$\delalpha$.  One can include these uncertainties,
$\sigma_q(j)$, in the fit and obtain a more realistic minimum
error estimate.  In this case, rather than a simple linear least-squares fit
one must use a method that allows errors in both the ordinate and the abscissa.
Since the uncertainties in $q$ are theoretical estimates and not statistical 
errors, this method is not technically completely consistent, but it should 
give a reasonable idea of the size of the effect.  As we find below,
the smallness of the claimed values for $\delalpha$ imply that the
uncertainties in $q$-values are not very important.

Table~\ref{tab:lineinfo} shows the minimum values 
of $\sigma_v$ for the portion of the combined 2004 spectra
containing each of the transitions that have a calculated q-value.
These results were calculated using co-added spectra from the standard 
XIDL HIRES pipeline that did 
not include the iodine cell wavelength recalibration; this 
simplifies the calculation but has little effect on these minimum error results.

The values range from 25 m/s for the strong Fe II $\lambda$1608 line to
153 m/s for the weak lines such as Fe II $\lambda$1611.
It is interesting to note that the maximum velocity precisions obtainable
with our data seem to be sufficient to measure the shifts predicted
by a changing value of $\delalpha$ at the level claimed by Murphy et al. 
(2003).  However, we also see that the weakness of the Fe II $\lambda$1611 line
means that the precision obtained from using only the Fe II 
$\lambda$1608/$\lambda$1611 pair may not be good enough for this purpose.

One can combine all these minimum $\sigma_v$ values using 
Equation~\ref{eqn:eqsigmav} to get
an overall minimum velocity error, but since $\delalpha$ is determined
by differences in redshifts this is not appropriate.
If one just wanted to determine how accurately the redshift of the entire
system could be determined, then one could combine the minimum errors
in Table~\ref{tab:lineinfo} to find $\sigma_v(min)({\rm all}) = 13$ m/s.
If we included only the lines for which we have iodine cell calibration
the result is $\sigma_v(min)({\rm calib}) = 15$ m/s.  
The 2002 data alone would give 38 m/s precision, which if combined with the
2004 data would give an ultimate velocity precision of 12 m/s.
Of course, as discussed below there are several important systematic errors 
that greatly increase these uncertainties.

Next we perform the least squares fit to find minimum possible errors 
on $\delalpha$ from this data
and display the results in Table~\ref{tab:minsigdelalpha}.
These uncertainties, e.g.  $\sigma(\delalpha) \geq 1.2 \ten{-6}$,
are reasonably competitive with errors 
quoted by Murphy, et al. (2001a; 2003; 2004),
with Chand, et al. (2004), and with Levshakov, et al (2006).
We again note that the latter two groups seem to have produced measurements of
$\delalpha$ with errors smaller than their minimum possible errors, something 
Murphy, Webb, and Flambaum (2006) have criticized, and which Levshakov et al.
subsequently corrected (Molaro, et al. 2008).

We note that minimum possible error coming from analysis of the Fe II 
$\lambda$1808/$\lambda$1611
pair is $\sigma(\delalpha) \geq 6.2\ten{-6}$, substantially less precise 
than could be found using
several lines.  As mentioned this is mostly because 
the Fe II $\lambda$1611 line is so weak (due to its rather low 
oscillator strength) and thus its redshift 
cannot be measured very
precisely.  Therefore, our original idea of using just these two lines is
probably not that useful.  In addition, examination of the lines shows that
it is only the saturated regions of the Fe II $\lambda$1608 line profile 
that are 
detectable in the Fe II $\lambda$1611 line profile.  Thus a joint $\chi^2$ fit
cannot accurately recover their relative velocity offset.  It seems it
is better to compare strong lines with other strong lines,
and weak lines with other weak lines.  We do this in the following section.

We can also test the importance of the uncertainties in $q$ listed 
in Table~\ref{tab:lineinfo},
by doing a linear fit that allows errors in both the abscissa and ordinate.
In this case, the resulting minimum error depends upon the fit value
of $\delalpha$, that is, the slope of the line given in Equation~\ref{eqn:eqvj}.
This is to be expected since 
if $\delalpha=0$, it does not matter what the values of
$x$ are, while if there is a large slope then uncertainty in $x$ will propagate
to uncertainty in $v$, and therefore uncertainty in $\delalpha$.
A simple way to estimate the increase in $\delalpha$ uncertainty is to just
do this propagation of errors, that is, change:
\begin{equation}
\sigma_v^2 \rightarrow \sigma_v^2 + \left(\delalpha\right)^2 \sigma_x^2,
\label{eqn:sigmaq}
\end{equation}
where $\sigma_x = -2 q c \lambda_0 \sigma_q$.
The results of this error propagation are also shown in 
Table~\ref{tab:minsigdelalpha} for the Murphy, et al. (2003) value 
$\delalpha = -5.4\ten{-6}$.  We note that because the
slope $\delalpha$ is small, 
the increase in uncertainty is
also quite small, no larger than around 14\% depending on the set of lines
used.  Thus previous workers who ignored this effect are not making a large
underestimation of their $\delalpha$ uncertainties and limits.
Because of the smallness of this effect, we will not consider the
uncertainty in $q$ in what follows.


It is important to realize that the minimum possible errors discussed above
are never reached in a real observation for several reasons as follows:

1. As extensively discussed in Murphy, et al. (2008, section 2.2.2), comparison
between transitions with different degrees of saturation will give velocity
uncertainties derived from Voigt profile fitting
substantially greater than the Fisher matrix minimum.  
This is because
velocity precision coming from the sharp edges of the saturated profiles
cannot be directly compared with the central regions of unsaturated profiles.
Especially in multicomponent systems, this will result in 
degeneracy between the component fit parameters, and  
make the redshift determination of any one component less accurate.
This is evident when the covariance matrix for one of our 5 component fits is
examined.  We see strong correlation between the redshifts of the
various components and between these redshifts and various other parameters 
such as the b-values.

We can investigate this effect directly in our data by comparing the 
formal parameter
fit errors with the minimum errors calculated above.
For example, a 5 component fit of the saturated Fe II $\lambda$1608 line
using the co-added data
gives a minimum possible redshift error of 33.4 m/s, while the fitting
program returns an error of 184 m/s on the strongest component.  

For Si II $\lambda$1808, an unsaturated line, the effect is smaller, but still pronounced:
32.6 m/s minimum error vs. 55 m/s fit result.  We note that the minimum
errors found from using formulas~\ref{eqn:eqsigmavi} and
\ref{eqn:eqsigmav} on the individual exposures are quite 
consistent with the minimum errors found from the co-added data:
For 3-component fit to Fe II 1$\lambda$1608 we find 32.6 m/s for co-added data, 
vs. 27.4 m/s from combining all 7 non-iodine cell exposures.
For Si II $\lambda$1808 the comparison is
50.4 m/s from the co-added compared to 49.9 m/s for individual exposures.
Note that errors returned from the fit software 
are actually also overly optimistic
because:

2. as discussed above there are other sources of error such as wavelength 
calibration, 

3. the errors on the flux are not really Gaussian which means that larger 
than Gaussian deviations occur, 

4. it is very difficult to determine the number of components needed to
fit each transition, especially for saturated, or nearly saturated lines.
The formal fit error returned on a component redshift does not include the
possibility of another component, which may be apparent in some transitions
and not in others.  The $\chi^2$ surface can have several local minima, and
small measurement errors can switch one between needing another component
and not needing another component.  Since it is the difference between component
redshifts that is used to determine $\delalpha$, this can be a source of
systematic error.  For example, we found examples where a region of
a spectrum was formally slighter better fit by two nearby components
than by a single component, but the
redshift of the components in the two cases shifted by around 1500 m/s, a 
value large compared to what is expected from $\delalpha$.
Murphy et al. (2008) investigated this effect in detail (see their Figure 7)
and found that while this effect does exist if the the number of fit 
components is less than the true number of components, this systematic
error is greatly reduced if the number of fit components is equal to or
greater than the true number of components.

\subsection{Fitting for $\delalpha$: simple method}
\begin{figure}
\epsscale{1.0}
\plotone{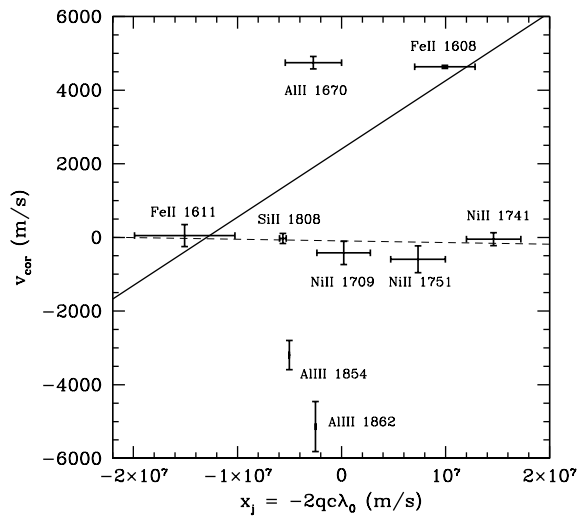}
\caption{
Linear fit simple method to find $\delalpha$ for the strongest component of
9 calibratable lines and non-iodine spectra.  
The solid line is the best fit for all 9 lines, while the dashed line is the
best fit for the
5 singly ionized unsaturated lines.
The slope of the 9 line fit is
$\delalpha = (186  \pm 6.3) \ten{-6}$, 
and the slope of the 5 line fit is $\delalpha = (-4.5  \pm 9.3) \ten{-6}$.
\label{fig:delalphanew9}
}
\end{figure}

We tried several methods of deriving a value of $\delalpha$ from our Voigt
profile fits.  
The simplest method is to fit each of our N lines
independently, and then do a least squares linear fit of 
Equation~\ref{eqn:eqvj}.
The value of $v_j$ can be found from 
\begin{equation}
v_j/c = (z_{\rm ave} - z_j )/(1+z_{\rm ave}),
\end{equation}
where $z_{\rm ave}$ is the redshift obtained by averaging over all
the transitions detected in a given velocity component. 
%
%
%
The result of this simple direct method for component 1 (the component 
with the largest column density) 
of the 9 lines for which we have both q-values and an iodine spectrum is shown
as the solid line in Figure~\ref{fig:delalphanew9}.
Here we plot $x_j = -2 q_j c \lambda_{0j}$ vs. the wavelength corrected
velocity $\vcor$ for each line, as well as the best linear fit
for the 9 lines (solid line) and the best linear fit for the 5 unsaturated,
lines that are singly ionized (dashed line).
For component 1 of the 9 lines, the fit slope is 
$\delalpha({\rm component\ }1)= (186\pm 6.3) \ten{-6}$, with
$\chidof=264$, formally a very strong detection, inconsistent with
other measurements.  For component 1 and the 5 lines, 
the fit slope is 
$\delalpha({\rm component\ }1)= (-4.5\pm 9.3) \ten{-6}$, with
$\chidof=1.09$, formally a nice null result consistent with, but much
weaker than results in the literature, and a factor of 3 worse than the
minimum possible error listed in Table~\ref{tab:minsigdelalpha}.  

However, we do not believe either of these results.
If a changing fine structure were the explanation 
for the variation in values of $\vcor$ in the 9 line fit, then
the points in Figure~\ref{fig:delalphanew9} would lie on a straight 
line within errors.
We see from the figure and from the terrible $\chidof$ that this 
is not the case.  
When the goodness of fit to a model is very bad, it means that
either the model is incorrect or that there are systematic errors
not included in the data uncertainties.  In such cases
the errors derived from the fit cannot be trusted, and
a better way to estimate the uncertainty 
is to increase the errors on the data points until $\chidof \approx 1$.
Doing this for the 9 line fit gives $\delalpha({\rm component\ }1) = 
(186 \pm 102)\ten{-6}$ ($\chidof=1.01$), 
a null result with a more realistic error estimate, 
but far from the sensitivity we expected from our data.

Another way to see that the first result above is meaningless
is to do the fit including the errors in $q$.  This gives an entirely
different answer which also is a terrible fit.   
However, when the errors are scaled as above,
this fit then gives nearly the same result as above but with larger errors.

The problem with the 9 line fit seems to be the saturated lines
Fe II $\lambda$1608 and Al II $\lambda$1670, as well as the aluminum lines
Al III $\lambda$1854 and $\lambda$1862. 
Since Al III is in a different ionization state than the other
ions, it might exist at a different physical location, and so maybe should
be left out of the fits.  
We want to discuss in more detail the saturated lines that are far from the 
others, but before doing that we note that this pattern repeats also in 
the $x$ vs. $\vcor$ fits of the 2nd strongest
component, the results of which are given in Table~\ref{tab:delalpha}. 

The fit redshifts of the first components of the
Fe II $\lambda$1608 and Al II $\lambda$1670 lines
differ from the dashed 5 line fit by about $\Delta z \sim 5\ten{-5}$.  
Some insight into possible reasons for this can be gained by noting that 
if we force this component of Fe II $\lambda$1608 to be the same as the average of the unsaturated transitions, we also get a completely satisfactory fit
to the Voigt profile:  $\chidof = 1487.7/1495=.995$ for the free fit, 
and $\chidof = 1503.4/1496=1.004$ ($Q=0.43$) for the 3 component
fit with one redshift forced to the average value.  
The reported formal fit errors on these redshifts are
$\sigma_z= 4.4\ten{-7}$, substantially smaller than the difference between 
the free fit and the forced fit.
Thus we see another possible source of error in determining $\delalpha$.  
The formal fit errors seem to underestimate the true range of acceptable 
redshift values, and therefore overestimate the precision with which 
$\delalpha$ can be determined by this method.  We note that when we force the
redshift of the first component of
Fe II $\lambda$1608 to agree with the other
lines, the 2nd and 3rd component also come into alignment.
Thus there is a degeneracy where the redshift and b-value of one component 
can play off the redshift and b-values of other components.
Another way this can be seen, is from the fact that the value derived 
for $\delalpha$ depends greatly on the error reported for the 
Fe II $\lambda$1608 redshift.
Figure~\ref{fig:delalphanew9} shows that the linear 
fit for $\delalpha$ is driven by the tiny error on the Fe II $\lambda$1608 
redshift.  Increasing that error by a factor of ten (to roughly equal the 
errors on the other redshifts) changes the value to
$\delalpha = (-9.3  \pm 9.0) \ten{-6}$.

Examination of the Al II $\lambda$1670 line shows a similar problem; again
the line is saturated and in this case there are certainly more than 3 
detectable components.  We are only allowing 3 total components so that 
we can compare component to component across other transitions where
only 3 components are detectable, but the fitting program can find different 
nearly equivalent ways to fit the line to 3 components since saturation
means less shape information is available.  

While the effect is largest for the saturated lines, we find the same 
effect for the Al III $\lambda$1864 and Al II $\lambda$1862, unsaturated
lines, where there are only 3 detectable components.  
Here again while the best fit redshift is far from the others, 
we can get a completely acceptable fit to the Voigt profile ($\chidof <1$) 
even while forcing the first component to the average.  
Again, when one component is forced to agree, the other components 
then also agree.

Thus, we note a systematic error that can give trouble in this type of fitting.
When a system has two (or more) components nearby in redshift space, 
there can be a near degeneracy where the two redshifts can play off each other,
moving significantly at the level of precision we are after, but yet
still giving very good Voigt profile fits.

Note that a linear fit to the 5 line set that includes the q errors
gives nearly the same answer as shown by the dashed line in
Figure~\ref{fig:delalphanew9}.
This is to be expected since the slope of the line, $\delalpha$, is small. 

We can repeat the above analysis for the 2nd and 3rd strongest components.
The results are given in Table~\ref{tab:delalpha} for various sets of lines.  
In these cases we mostly find null results such as 
$\delalpha({\rm component\ }2) = (14 \pm 12)\ten{-6}$ with $\chidof=92$
for all nine lines.  This is because the lower column densities mean
larger formal fit errors.

There are additional components detectable in some of the lines, 
but they are much weaker and do not contribute much. 
To get a final answer using this method one could average the above results,
which actually gives a result with larger formal error than from the first 
component alone, again showing that systematic errors are dominating 
this procedure.

\subsection{Fitting for $\delalpha$: many parameter joint fit}
The above method, while useful for showing errors arising from
fitting degeneracies, is not
really correct because for it to work one
must match up the various components fit in different lines
in order to properly compare their redshifts.
As noted, these fitting degeneracies allow the redshifts of
two nearby components to be traded against each other and
against the b-values, resulting in  
correlations that cause the uncertainty in the component redshifts
to be larger than the formal fit errors.

More properly, one should do
a large joint fit for all system components simultaneously, that is
tie together the different velocity components.
This is standard practice and is used by Murphy et al., Chand, et al. Levshakov,
et al., etc.
Thus, for example, the fit to the 9 lines above
by the previous method used a total of 81 parameters: redshift, column
density and b-value for each of 3 components of each of 9 lines.
A global joint fit might
might have 9 parameters for the redshift, column density, and b-value
of each of 3 components, plus some column density offsets to allow for 
different elemental abundances 
(1 parameter per component per additional element or 
12 including Fe II, Ni II, Si II, and Al II and Al III).  
An additional b-value offset is
probably necessary for each component of the Al III lines since these are 
in a different ionization state (3 more parameters).
Then adding $\delalpha$ as a final parameter, there would be
a total of 25 parameters.

The results of such a global joint fit method
are shown
in Table~\ref{tab:delalpha} for various sets of lines.  
Table~\ref{tab:delalpha} shows that
results we find for $\delalpha$ vary over a wide range of values,
from significant detection $\delalpha=(-26 \pm 8) \ten{-6}$ for the 
Fe II $\lambda$1608/Fe II $\lambda1611$ pair, to
to a null result $\delalpha= (-0.07 \pm 3.9) \ten{-6}$ for the nine-line 
fit.
Looking over all the results in Table~\ref{tab:delalpha} we see that
both the method used and the lines selected can make a significant difference
in the final result.  If our errors were under control,
this should not be the case.
Note that in most cases, the errors for the global fit method 
method hover around 
$4 \ten{-6}$; roughly half the errors of around $8 \ten{-6}$
found in the simple method described in the previous section.
In Table~\ref{tab:fitcomponents} we display the Voigt profile components that 
resulted from one of our joint global fits.

\subsection{Fitting for $\delalpha$: VPFIT global joint fit}
One might worry that the puzzling results we find come from errors in
our fitting program, so as a final check, we redid our calculation
of $\delalpha$ using the new version of VPFIT (Carswell, et al. 2008) which 
includes the possibility of doing a global joint fit for $\delalpha$.
These results are also shown in Table~\ref{tab:delalpha}.  For the
6 line set (Fe II $\lambda$1611/$\lambda$1608, Ni II$\lambda$1709/$\lambda$1741/$\lambda$1751, 
Si II$\lambda$1808) we studied
above we find $\delalpha = (-9.1 \pm 7.7) \ten{-6}$ 
($\chidof=3.1, Q=10^{-139}$), 
while for the set of all 16 lines we find
find $\delalpha=(-17\pm10)\ten{-6}$
($\chidof=6.0, Q=0$). 
Both these results are 
within the range found by our different fitting methods. 

\section{Discussion}
We tried and failed to give a definitive answer to the question of whether
the fine structure constant was different at early times in high redshift
Ly alpha systems.  
In order to investigate this problem in detail we
used data taken through the Keck iodine cell and wrote our fitting 
software from scratch.  
Using the iodine cell for wavelength calibration we found a
serious source of systematic error
that did not allow calibration of the Keck HIRES spectrograph to the
precision needed.  
We also found degeneracies in the fitting procedures
that added to the calibration systematic errors.

Due to all the systematic errors we were able to derive various results,
running from very significant detections, to strong null limits.
Does this imply that a meaningful measurement of or limit on $\delalpha$ is
impossible using Keck HIRES?  It is not clear.  Perhaps more careful
attention to Voigt profile fitting, as advocated by M.~Murphy (private
communication 2008) will solve the fit degeneracy problem.  
Perhaps a careful selection of absorbers would also help, or 
trying to focus on systems with a single component.
Perhaps the wavelength calibration errors we discovered
can be corrected, or perhaps they will average away
if a large sample of absorbers is considered.  
In this paper we raise these questions, but do not answer them.

We also used a Fisher matrix technique to investigate 
the minimum possible errors in $\delalpha$ that our spectra's signal/noise
would allow.  We found our fit results did not exceed these limits.    For example,
for the set of 9 lines used in most of our analyses, the minimum
possible error on $\delalpha$, as given in Table~\ref{tab:minsigdelalpha},
is $2.52\ten{-6}$, a result consistent with our fit results.  
Since the calculation of minimum possible errors is not difficult, we suggest
that workers always calculate them and never report results with uncertainties
smaller than the data theoretically can allow. 

While we have not yet looked carefully at data or analysis done by other
workers in the field, we worry that some of the systematic errors and
overestimation of precision we found here may also be present
in other analyses.  
Thus one possible explanation for the discrepant findings on $\delalpha$
discussed in the introduction, is that several workers in the field 
are overestimating the precision of their measurements and the discrepancies 
reported in the literature are due to random fluctuations occurring within the 
larger, under-reported, systematic errors.

At this point it is not clear how to make further progress in this subject
using Keck HIRES, but other techniques such as frequency combs 
(Steinmetz, et al. 2008) may become available and be of use in resolving the
question.  In addition, proposed new instruments (e.g., CODEX for E-ELT or
ESPRESSO for the VLT) are being designed for Doppler measurement
stability and will hopefully be be free of these problems.

\acknowledgments
We thank David Kirkman, Bob Carswell, Marc Rafelski, Joel Heinrich, 
John Johnson, and Michael J. Simmonds for helpful discussions.
We especially thank Michael Murphy for many insightful comments,
suggestions and questions, including several that led to corrections of 
errors in early versions of this paper.  
Finally, we thank Nao Suzuki for discussion of an early version of this
paper where we discovered that he had already understood and published
several of the points made here.

K.G. and J.B.W. were supported in part by the DoE under grant
DE-FG03-97ER40546.  J.X.P. is partially supported by an NSF CAREER grant
(AST-0548180), and J.X.P. and A.M.W. are supported in part by
NSF grant (AST-07-09235).
The W. M. Keck Observatory is operated as a scientific partnership 
among the California Institute of Technology, the University of California 
and the National Aeronautics and
Space Administration.  The Observatory was made possible by the generous
financial support of the W. M. Keck Foundation.  The 
authors wish to recognize and
acknowledge the very significant cultural role and reverence that the summit
of Mauna Kea has always had within the indigenous Hawaiian community.  We are
most fortunate to have the opportunity to conduct observations from this
mountain.

\clearpage

\begin{deluxetable}{cccccccccc}
\tabletypesize{\scriptsize}
\tablecaption{Journal of PHL957 Observations
\label{tab:journal}
}
\tablewidth{0pt}
\tablehead{
\colhead{Date} & \colhead{ID} & \colhead{Iodine Cell?} 
&\colhead{Time (UT)} &\colhead{Exposure (s)} 
&\colhead{Tempin$^a$ ($^0$C)}
&\colhead{Arc ID}
&\colhead{Arc Time (UT)}
&\colhead{Moved?$^b$}
&\colhead{$\Delta$ Tempin ($^0$C)}
}
\startdata
1 nov 02 & 33 & out & 5:26 & 1800 & 3.75  &&&&\\
1 nov 02 & 34 & out & 5:58 & 1800 & 3.76 &&&&\\
1 nov 02 & 35 & in & 6:30 & 1800 & 3.79 &&&&\\
1 nov 02 & 36 & in & 7:01 & 1800 & 3.76 &&&&\\
1 nov 02 & 47 & out & 7:58 & 1800 & 3.61 &&&&\\
1 nov 02 & 48 & out & 8:30 & 1800 & 3.68 &&&&\\
1 nov 02 & 49 & in & 9:03 & 1800 & 3.68 &&&&\\
1 nov 02 & 50 & in & 9:34 & 1800 & 3.78 &&&&\\
1 nov 02 & 60 & out & 10:34 & 1800 & 3.71 &&&&\\
1 nov 02 & 61 & out & 11:06 & 1800 & 3.75 &&&&\\
1 nov 02 & 62 & in & 11:39 & 1800 & 3.71 &&&&\\
3 oct 04 & 67/3-0 & out &9:18  & 3600 & 4.26 & 66 & 09:15 &no & 0.042\\
3 oct 04 & 68/3-1 & in & 10:31 & 3600 & 4.21 & 66 & 09:15 & no & 0.097\\
3 oct 04 & 69/3-2 & in & 11:33 & 3600 & 4.07 & 66 & 09:15 & no & 0.236\\
3 oct 04 & 70/3-3 & out & 12:35 & 3600 & 4.00 & 66& 09:15 &no& .0306\\
4 oct 04 & 1096/4-0 & out & 9:25 & 3600 & 3.00 & 1144 & 15:35 &yes  & -0.153\\
4 oct 04 & 1098/4-1 & in & 10:56 & 3600 & 2.97 & 1144 & 16:35 & yes & -0.125\\
4 oct 04 & 1099/4-2 & in & 11:57 & 3600 & 3.01 & 1144 & 16:35 & yes & -0.167\\
4 oct 04 & 1100/4-3 & out & 12:58 & 3600 & 2.93 &1144 & 16:35 & yes& -0.083\\
5 oct 04 & 2094/5-0 & in & 8:25 & 3600 & 2.87 & 2026 & 3:01 & yes  & 0.125\\
5 oct 04 & 2095/5-1 & out & 9:27 & 3600 & 2.86 & 2107& 15:46& yes& 0.430\\
5 oct 04 & 2096/5-2 & out & 10:29 & 3600 & 2.91 &2107& 15:46& yes& 0.375\\
5 oct 04 & 2097/5-3 & in & 11:30 & 3600 & 3.62 & 2107$^*$ & 15:46 & yes & -0.333\\
5 oct 04 & 2098/5-4 & out & 12:32 & 2700 & 3.55 &2107& 15:46 &yes& -0.264\\

\enddata
\tablenotetext{a}{Temperature inside the the HIRES enclosure.}
\tablenotetext{b}{Whether or not the grating was moved between Th/Ar arc and 
data exposures}
\tablenotetext{*}{
This exposure was also calibrated with Th Ar arc 2026;  see text}
\end{deluxetable}

\begin{deluxetable}{cccccc}
\tabletypesize{\scriptsize}
\tablecaption{Line Information \label{tab:lineinfo}}
\tablewidth{0pt}
\tablehead{
\colhead{Transition } &
\colhead{Echelle order } &
\colhead{q value (cm$^{-1}$)$^{\rm a}$ } &
\colhead{ min $\sigma_v$ (m/s) (2004 data)}  &
\colhead{Iodine cell coverage? } }
\startdata
Fe II $\lambda$1608.45 & 67 & $-1030 \pm 300^*$ & 25.0 & yes  \\
Fe II $\lambda$1611.20 & 67 &  $1560 \pm 500^*$ & 153 & yes  \\
Al II $\lambda$1670.79& 65 & $270 \pm  ?^\ddagger$ & 34.0 & yes  \\
Ni II $\lambda$1709.60 & 63 & $-20 \pm 250^{**}$ & 83.1 & yes \\
Ni II $\lambda$1741.55 & 62 & $-1400 \pm 250^{**}$ & 48.7 & yes \\
Ni II $\lambda$1751.92 & 62 & $-700 \pm 250^{**}$ & 70.8 & yes \\
Si II $\lambda$1808.01 & 60 & $520 \pm 30^{**}$ & 36.4 & yes \\
Al III $\lambda$1854.72 & 58 & $458 \pm  2^{***}$ & 76.0 & yes \\
Al III $\lambda$1862.79 & 58 & $224 \pm  1^{***}$ & 125 & yes \\
\tableline
Si II $\lambda$1526.71 & 71 & $50 \pm 30^{**}$ & 28.8 & no \\
Zn II $\lambda$2026.14 & 53 & $2488 \pm 25^\dagger$ & 129 & no  \\
Zn II $\lambda$2062.66 & 52 & $1585 \pm 25^\dagger$ & 229 & no  \\
Cr II $\lambda$2056.26 & 52 & $-1030 \pm 150^{**}$ & 89.9 & no  \\
Cr II $\lambda$2062.24 & 52 & $-1168 \pm 150^{**}$ & 102 & no  \\
Cr II $\lambda$2066.16 & 52 & $-1360 \pm 150^{**}$ & 143 & no  \\
Fe II $\lambda$2344.21 & 46 & $1540 \pm 400^{*}$   & 41.7 & no  \\
\enddata
\tablenotetext{a}{
q-values marked $^*$ are from Dzuba, et al (2001);
marked $^{**}$ from Porsev, et al (2007); marked $^{\ddagger}$ from Murphy, et al. 2001a; 
marked $^\dagger$ from Savukov \& Dzuba (2007); marked $^{***}$ from Dzuba \& Flambaum (2008).}
\end{deluxetable}


\begin{deluxetable}{cccc}
\tablecaption{ Minimum possible errors in $\delalpha$ 
\label{tab:minsigdelalpha}}

\tablewidth{0pt}
\tabletypesize{\scriptsize}
\tablehead{
\colhead{Line set} &
\colhead{$\delalpha$(min) (no error in q)} &
\colhead{$\delalpha$(min) (including error in q)} &
\colhead{$\delalpha$(min) (no error in q; w/o I2 exposures)}} 
\startdata
All 16 lines (2002 and 2004 data) & $1.17\ten{-6}$ & $1.32\ten{-6}$ & -- \\
All 16 lines  & $1.21\ten{-6}$ & $1.36\ten{-6}$ & $1.50\ten{-6}$\\
9 calibratable lines & $2.00\ten{-6}$ & $2.14\ten{-6}$ & $2.52\ten{-6}$\\
7 calibratable, unsaturated lines & $2.69\ten{-6}$ & $2.77\ten{-6}$ & $3.28\ten{-6}$\\
Fe II $\lambda$1608/$\lambda$1611 pair  & $6.18\ten{-6}$ & $6.31\ten{-6}$ & $6.20\ten{-6}$\\
\enddata
\tablecomments{Values are for 2004 data only and include exposures with
and without the iodine cell in place, except where noted.}
\end{deluxetable}


\begin{deluxetable}{ccrrc}
\tabletypesize{\scriptsize}
\tablecaption{Fit results for $\delalpha$
\label{tab:delalpha}}
\tablewidth{0pt}
\tablehead{
\colhead{Method} &
\colhead{Line set} &
\colhead{$\delalpha /\aten{-6}$} &
\colhead{$\chidof$} & 
\colhead{probability ($Q$)}}
\startdata
Simple (comp 1) & the (5) 	& $-4.5 \pm 9.3$	 & 1.09& 0.35 \\
Simple (comp 1)& (5)+Fe II $\lambda$1608 	& $213 \pm 7.1$ & 323& $\aten{-278}$ \\
Simple (comp 1)& (5)+Al III $\lambda$1854/$\lambda$1862  & $6.7\pm9.2$ & 22 & $2\ten{-22}$ \\
Simple (comp 1)& (5)+Fe II $\lambda$1608 +Al III $\lambda$1854/$\lambda$1862	& $237 \pm 6.8$ 
	& 248& $\aten{-318}$ \\
Simple (comp 1)& the (9) & $186 \pm 6.3$  
	& 264& 0 \\
Simple (comp 2) & the (5) 	& $-7.1 \pm 12$ & 0.66& 0.57 \\
Simple (comp 2)& (5)+Fe II $\lambda$1608 	& $15 \pm 12$ & 75& $\aten{-63}$ \\
Simple (comp 2)& the (9) &  $14 \pm 12$	 & 92& $\aten{-136}$ \\
Simple (comp 3)& (5)+Fe II $\lambda$1608 	& $25 \pm 51$ & 3.8& 0.004 \\
\tableline
global joint fit & Fe II $\lambda$1608 and Fe II $\lambda$1611 pair & $-26\pm8.1$ & 0.96& - \\
global joint fit & the (5) & $-3.3\pm4.3$ & 0.80& - \\
global joint fit & (5) + Fe II $\lambda$1608 & $-18 \pm 3.8$ & 0.92& - \\
global joint fit & (5) + Al III $\lambda$1854/$\lambda$1862 & $-6.3\pm4.3$ & 0.83& - \\
global joint fit & the (9)  & $-0.70 \pm 3.9$ & 1.01& -\\
\tableline
VPFIT global (w/o iodine) & (5) + Fe II $\lambda$1608 &  $-9.1\pm7.7$ & 3.1 &$10^{-139}$ \\
VPFIT global (w/o iodine) & all 16 & $-17\pm10$ & 6.0 & $0$\\
\enddata
\tablecomments{
The five (5) lines included in almost all the fits are 
Fe II $\lambda$1611, Ni II $\lambda$1709/$\lambda$1741/$\lambda$1751, and Si
II $\lambda$1808. The nine (9) lines are the (5) plus 
Fe II $\lambda$1608, 
Al III $\lambda$1854/$\lambda$1862 and Al II $\lambda$1670. 
The VPFIT spectra were handled differently and did not include the iodine wavelength correction, so VPFIT 
results are not expected to be the same as our joint global fit results.
}
\end{deluxetable}

\begin{deluxetable}{cccc}
\tabletypesize{\scriptsize}
\tablecaption{Component fit values for a 3 component joint global fit of 9 
lines. 
\label{tab:fitcomponents}}
\tablewidth{0pt}
\tablehead{
\colhead{Component} &
\colhead{redshift $z$} &
\colhead{velocity width $b$ (km/s)} &
\colhead{log column density $\log(N)$}}
\startdata
Comp 1 &  $2.3090291 \pm 9.58\ten{-7}$ & $6.41 \pm 0.05$ & $14.76 \pm 0.029$ \\
Comp 2 &  $2.3091517 \pm 1.68\ten{-6}$ & $7.75 \pm 0.12$ & $14.563 \pm 0.035$ \\
Comp 3 &  $2.3094793 \pm 9.49\ten{-7}$ & $16.44 \pm 0.11$ & $14.039 \pm 0.021$\\
\enddata
\tablecomments{
The 9 lines included in the fit are the ones listed in the Introduction.
Each component of each element has an additional fit parameter (not listed)
which is a column density offset to allow for different elemental abundances.
This fit resulted in a fit value of $\delalpha = (-0.070 \pm 3.9) \ten{-6}$ 
and $\chidof = 14138/13891 = 1.01$.
}
\end{deluxetable}

\end{document}